\documentclass[twocolumn,superscriptaddress,aps,preprintnumbers,amsmath,amssymb,prl,nofootinbib]{revtex4}

\usepackage{epsfig}
\usepackage{amsmath}
\usepackage{amssymb}
\usepackage{subfigure}
\usepackage{cancel}
\usepackage{textcomp}
\usepackage{calc}
\usepackage{graphicx}
\usepackage{dcolumn}
\usepackage{color}

\usepackage{hyperref}
\hypersetup{colorlinks=true,urlcolor=blue,linkcolor=red,citecolor=green}

\begin{document}


\title{Unified Model of D-Term Inflation}

\author{Valerie Domcke}
\affiliation{APC / PCCP, Paris Diderot University, 75013 Paris, France}

\author{Kai Schmitz}
\affiliation{Max-Planck-Institut f\"ur Kernphysik (MPIK), 69117 Heidelberg, Germany}


\begin{abstract}
Hybrid inflation, driven by a Fayet-Iliopoulos (FI) D term, is an
intriguing inflationary model.
In its usual formulation, it however suffers from several shortcomings.
These pertain to
the origin of the FI mass scale,
the stability of scalar fields during inflation,
gravitational corrections in supergravity, as well as to
the latest constraints from the cosmic microwave background.
We demonstrate that these issues can be remedied
if D-term inflation is realized in the context of strongly
coupled supersymmetric gauge theories.
We suppose that the D term is generated
in consequence of dynamical supersymmetry breaking.
Moreover, we assume canonical kinetic terms in the Jordan frame as well as 
an approximate shift symmetry along the inflaton direction.
This provides us with a unified picture of D-term
inflation and high-scale supersymmetry breaking.
The D term may be associated with a gauged
$U(1)_{B-L}$, so that the end of inflation spontaneously breaks
$B$$-$$L$ in the visible sector.
\end{abstract}


\date{\today}
\maketitle


Cosmic inflation~\cite{Guth:1980zm} is a successful paradigm in our understanding
of the early universe.
It is, however, still unclear how to correctly
embed inflation into particle physics~\cite{Lyth:1998xn}.
One promising ansatz is the idea of hybrid inflation~\cite{Linde:1991km},
which establishes a connection between inflation and grand unification.
Hybrid inflation exits into the subsequent radiation-dominated phase
via a \textit{waterfall transition}.
In the context of a given grand unified theory (GUT), this phase
transition may then be identified with the spontaneous breakdown of a local GUT symmetry.


Depending on the type of GUT symmetry, the waterfall transition
may have important consequences for the particles of the standard model (SM).
Here, a prominent example is the spontaneous breaking
of $U(1)_{B-L}$~\cite{Buchmuller:2010yy}, i.e., the Abelian gauge symmetry associated
with the difference between baryon number~$B$ and lepton number~$L$.
The end of inflation is then accompanied by
the generation of large $L$-violating Majorana masses for
a number of right-handed neutrinos, which sets the stage
for the type-I seesaw mechanism~\cite{Minkowski:1977sc}
as well as for baryogenesis via leptogenesis~\cite{Fukugita:1986hr}.
Hybrid inflation ending in a $B$$-$$L$ phase transition, thus,
promises to provide an appealing framework for the early universe
that not only determines the initial conditions of the hot thermal phase,
but which also explains the smallness of the SM neutrino masses.


In its simplest, nonsupersymmetric form, hybrid inflation
predicts the primordial scalar power spectrum to be blue-tilted, which is by now
observationally ruled out at a level of more than $5\,\sigma$~\cite{Ade:2015lrj}.
This problem can be avoided in supersymmetry (SUSY), where scalar
and fermion loops generate a logarithmic effective potential.
Supersymmetric hybrid inflation comes in two variants, depending on
whether the vacuum energy density during
inflation is dominated by a nonzero F~term~\cite{Copeland:1994vg} or
D~term~\cite{Binetruy:1996xj}.
In F-term hybrid inflation (FHI), the inflaton field itself has a large
F term during inflation. 
In combination with $R$ symmetry breaking, this
results in a dangerous supergravity (SUGRA) tadpole term~\cite{Buchmuller:2000zm},
which breaks the rotational invariance in field space, generates
a false vacuum at large field values, and potentially spoils slow-roll
inflation.
In D-term hybrid inflation (DHI), the superpotential of the inflationary
sector has, by contrast, zero vacuum expectation value (VEV) at all times,
so that SUGRA corrections tend to become more manageable.
Moreover, DHI is based on a nonzero Fayet-Iliopoulos (FI) D term~\cite{Fayet:1974jb}
and, hence, does not require a dimensionful input parameter in the superpotential.


In this paper, we shall construct a consistent SUGRA model in which
DHI is driven by the D term associated with a local $U(1)_{B-L}$ symmetry.
Despite the absence of the inflaton tadpole term,
this is still a difficult task for at least five reasons:
(i) The consistent embedding of the FI term into SUGRA is a subtle
issue that has been the subject of a long debate in the
literature.
On the one hand, constant, field-independent FI terms always require
an exact global symmetry~\cite{Komargodski:2009pc}, which conflicts
with the expectation that quantum gravity actually does not admit such
symmetries~\cite{Banks:2010zn}.
On the other hand, field-dependent FI terms
(such as those in string theory~\cite{Dine:1987xk})
imply the existence of a shift-symmetric modulus field~\cite{Komargodski:2010rb},
which causes cosmological problems~\cite{Coughlan:1983ci},
as long as it is not properly stabilized~\cite{Binetruy:2004hh}.
(ii) The sfermions in the minimal
supersymmetric standard model (MSSM) carry nonzero $B$$-$$L$ charges
and, thus, acquire D-term-induced masses during inflation~\cite{Babu:2015xba}.
Some of these masses are tachyonic and may, hence, destabilize
the corresponding directions in scalar field space~\cite{Domcke:2014zqa}.
(iii) General arguments in SUGRA~\cite{Dumitrescu:2010ca}
indicate that a nonzero D term is typically
accompanied by a comparatively larger F term, $\left|F\right| \gtrsim \left|D\right|$.
If SUSY breaking is mediated to the visible sector
by ordinary gravity mediation~\cite{Nilles:1983ge},
this implies that the inflaton picks up a gravity-mediated soft 
mass of the order of the gravitino mass, $m_{3/2} \sim \left|F\right|/M_{\rm Pl}$,
that necessarily exceeds the Hubble rate during inflation, $H \sim |D|/M_{\rm Pl}$.
DHI in combination with ordinary gravity mediation, therefore,
also faces the $\eta$ slow-roll problem in SUGRA~\cite{Dine:1983ys}.
(iv) In the global-SUSY limit, DHI predicts a scalar spectral index
of $n_s \simeq 0.98$~\cite{Binetruy:1996xj}.
This deviates from the latest
value reported by the PLANCK collaboration,
$n_s^{\rm obs} =0.9677\pm0.0060$~\cite{Ade:2015lrj}, by about $2\,\sigma$.
SUGRA corrections may help to reach
better agreement with the data~\cite{Seto:2005qg}.
But in general, realizing a spectral index of $n_s \simeq 0.96$ in DHI
is a nontrivial task.
(v) In its standard formulation, DHI is driven by a D term associated
with a $U(1)$ symmetry that becomes spontaneously broken only during
the waterfall transition at the end of inflation.
This results in the production of cosmic strings, which impact the 
scalar power spectrum of the cosmic microwave background (CMB). 
The recent CMB bounds on the tension of such cosmic strings~\cite{Ade:2013xla}
severely constrain the parameter space of standard DHI.


We now argue that all of these issues can be remedied
as soon as one makes the following three assumptions:
(i) The FI term is dynamically generated in the
context of dynamical SUSY breaking (DSB)~\cite{Domcke:2014zqa}.
(ii) DHI is embedded into Jordan-frame SUGRA with canonical
kinetic terms for all fields~\cite{Ferrara:2010yw}.
(iii) There exists an approximate shift symmetry in the direction
of the inflaton field~\cite{Kawasaki:2000yn}.
For a more comprehensive account of
our idea, see~\cite{Domcke:2017rzu}.


As far as the generation of the FI term is concerned, we follow
the discussion in~\cite{Domcke:2014zqa}.
We assume that SUSY is broken in a hidden sector by
the dynamics of a strongly interacting supersymmetric gauge theory.
To this end, we shall employ the
Izawa-Yanagida-Intriligator-Thomas (IYIT) model~\cite{Izawa:1996pk}
in its $SU(2)$ formulation, i.e., the simplest conceivable DSB model
with vector-like matter fields.
At high energies, the IYIT model consists of four quark fields $\Psi^i$
in the fundamental representation of $SU(2)$.
At energies below the dynamical scale $\Lambda_0$, these quarks
condense into six gauge-invariant meson fields,
$M^{ij} \simeq \left<\Psi^i\Psi^j\right>/\left(\eta^2\Lambda\right)$,
where $\Lambda \simeq \Lambda_0/\eta$ and $\eta \simeq 4\pi$~\cite{Manohar:1983md}.
The scalar mesons span a quantum moduli space of degenerate supersymmetric vacua,
subject to a particular constraint on their Pfaffian,
$\textrm{Pf}\left(M^{ij}\right) \simeq \Lambda^2$~\cite{Seiberg:1994bz}.
In order to break SUSY in this model, one couples the high-energy theory
to a set of six $SU(2)$ singlets, $Z_{ij}$, so as to lift the flat directions
in moduli space.
At high and low energies, the IYIT superpotential respectively reads as follows,
\begin{align}
W_{\rm hid}^{\rm HE} = \frac{1}{2} \lambda_{ij} Z_{ij} \Psi^i \Psi^j \:\:\rightarrow\:\:
W_{\rm hid}^{\rm LE} \simeq \frac{1}{2} \lambda_{ij} \Lambda Z_{ij} M^{ij} \,,
\end{align}
where $\lambda_{ij}$ is a matrix of Yukawa couplings.
SUSY is now broken \`a la O'Raifeartaigh~\cite{ORaifeartaigh:1975nky},
as the F-term conditions for the singlet fields $Z_{ij}$ are incompatible
with the Pfaffian constraint.
A crucial observation for our purposes is that the IYIT model exhibits
an axial $U(1)$ flavor symmetry associated with a quark phase rotation,
$\Psi^i\rightarrow e^{iq_i\alpha}\Psi^i$.
We shall now identify this symmetry with $U(1)_{B-L}$ and promote
it to a weakly gauged local symmetry.
In doing so, we suppose that two quarks carry charge $q_0/2$,
while the other two carry charge $-q_0/2$.
In this case, we have to deal with six mesons (and similarly six singlets)
with charges $q_0$, $-q_0$, and four times $0$, respectively.
Here, we assign the $B$$-$$L$ charges in such a way that
the charged mesons, $M_\pm$, have the smallest Yukawa couplings, $\lambda_\pm$.
During SUSY breaking, it is therefore the fields $M_\pm$
that acquire nonzero VEVs.
The neutral mesons and singlets remain stabilized
at their origin.
In the weakly gauged limit, one finds
(see~\cite{Domcke:2014zqa,Harigaya:2015soa,Babu:2015xba,Schmitz:2016kyr}
for more details on the dynamics of the IYIT model and its applications),
\begin{align}
\left<M_\pm\right> = \frac{\lambda}{\lambda_\pm} \Lambda \,, \quad
\lambda = \sqrt{\lambda_+\lambda_-} \,.
\end{align}
These VEVs break $B$$-$$L$ spontaneously, which
results in an effective FI term in the $B$$-$$L$ D-term scalar potential,
\begin{align}
V_D = \frac{g^2}{2}\left[q_0\left(\left|M_-\right|^2-
\left|M_+\right|^2\right) + \cdots\right]^2 \,.
\end{align}
Here, $g$ denotes the $B$$-$$L$ gauge coupling, while the ellipsis
stands for all further fields that are charged under $U(1)_{B-L}$.
One then obtains for the FI mass scale $\xi$,
\begin{align}
\xi = \left<M_-\right>^2 - \left<M_+\right>^2 =
\frac{2}{\rho^2}\left(1-\rho^4\right)^{1/2} \Lambda^2 \,,
\label{eq:xi}
\end{align}
where $\rho = \left[\left(\lambda_+/\lambda_-+\lambda_-/\lambda_+\right)/2\right]^{-1/2}$
is a measure for the degeneracy among the Yukawa couplings
$\lambda_+$ and $\lambda_-$.
For $\lambda_+ \simeq \lambda_-$, $\rho$ is close to unity;
for a strong hierarchy among $\lambda_+$ and $\lambda_-$, it takes
a value close to zero.
In the following, we will assume that $\lambda_+$ and $\lambda_-$ are both
of the same order of magnitude.
Averaging all possible values of $\rho$ under this assumption
(varying $\lambda_+$ and $\lambda_-$ on a linear scale)
then results in an expectation value of $\left<\rho\right> \simeq 0.80$.


The FI parameter in Eq.~\eqref{eq:xi} is a field-dependent FI term,
as it originates from the VEVs of the two meson fields $M_\pm$.
The modulus field associated with this FI term is nothing but
the $B$$-$$L$ Goldstone multiplet,
$A = \left(\left<M_+\right>M_+-\left<M_-\right>M_-\right)/f_a$, where
$f_a$ is the Goldstone decay constant,
$f_a^2 = \left<M_+\right>^2+\left<M_-\right>^2$.
The pseudoscalar in $A$ is absorbed by
the massive $B$$-$$L$ vector boson, while the real scalar
in $A$ is stabilized by an F-term-induced mass,
$m_F = \rho\lambda\Lambda$.
The same holds true for the fermion in $A$.
Owing to the fact that our FI term is generated in conjunction with
dynamical SUSY breaking, we therefore do not face any modulus problem.
Our model, thus, avoids the problems described
in~\cite{Komargodski:2009pc,Komargodski:2010rb}.


In the SUSY-breaking vacuum at low energies, the IYIT model effectively
reduces to the Polonyi model~\cite{Polonyi:1977pj}
with an effective superpotential of the following form~\cite{Schmitz:2016kyr},
\begin{align}
W_{\rm hid} = \mu^2\,X + w \,.
\label{eq:Polonyi}
\end{align}
Here, $\mu^4 = \lambda_+^2\left<M_+\right>^2 \Lambda^2 +
\lambda_-^2\left<M_-\right>^2 \Lambda^2 = 2\,\lambda^2\Lambda^4$ denotes the
F-term SUSY-breaking scale, $X = \left(Z_+ + Z_-\right)/\sqrt{2}$ is
the Polonyi field, and $w$ is an $R$-symmetry-breaking
constant that needs to be added to $W_{\rm hid}$, so as to achieve
zero cosmological constant in the true vacuum.


We now couple the effective Polonyi model in Eq.~\eqref{eq:Polonyi} to SUGRA.
In doing so, we shall work in Jordan-frame supergravity with
canonical kinetic terms~\cite{Ferrara:2010yw}. 
The total K\"ahler potential of our theory
is therefore given as
\begin{align}
K_{\rm tot} = -3 M_{\rm Pl}^2 \, \ln\left(-\frac{\Omega_{\rm tot}}{3 M_{\rm Pl}^2}\right) \,,
\end{align}
where $\Omega_{\rm tot} = -3 M_{\rm Pl}^2 + F_{\rm tot}$ is the frame function of
the Jordan frame.
We assume that the kinetic function $F_{\rm tot}$ can be split into
separate contributions from the hidden, visible, and inflaton sector.
Schematically, we may write
\begin{align}
F_{\rm tot} = F_{\rm hid} + F_{\rm vis} + F_{\rm inf} +
\frac{1}{M_*^2}\, F_{\rm hid}\, F_{\rm vis} \,,
\label{eq:F}
\end{align}
such that the inflaton
sector becomes sequestered from the hidden sector~\cite{Inoue:1991rk}.
This serves the purpose to protect the inflaton
from a SUGRA mass of the order of $m_{3/2} \gtrsim H$,
which would otherwise spoil slow-roll inflation.
Meanwhile, the MSSM sfermions do acquire soft masses via gravity mediation.
These may be much larger than $m_{3/2}$, provided that the
mass scale $M_*$ is parametrically smaller than
the reduced Planck mass~$M_{\rm Pl} \simeq 2.44\times10^{18}\,\textrm{GeV}$.
In particular, by choosing an appropriate value of $M_*$,
the MSSM sfermions are also sufficiently stabilized during inflation,
even if inflation is driven by a $B$$-$$L$ D term.


For $F_{\rm hid} = \left|X\right|^2 + \textrm{[other fields]}$, we obtain
for the Polonyi VEV during ($f\neq0$) and after ($f=0$) inflation,
\begin{align}
\left<X\right> = \frac{\left(k-4/3\right)^{1/2}}{\left[\left(1-f\right)k-4/3\right]k^{1/2}}
\frac{2 M_{\rm Pl}}{\sqrt{3}} \,.
\label{eq:XVEV}
\end{align}
Here, $f$ is related to the kinetic function of the inflaton field $S$,
$f = \left(F_{\rm inf} - \partial_S F_{\rm inf}\, \partial_{S^\dagger} F_{\rm inf}\right)/
\left(3M_{\rm Pl}^2\right)$.
$k$ is a ratio of different contributions to the total Polonyi mass,
\begin{align}
k = \left[\left(m_{1\ell}^J\right)^2 + 2\,H_J^2\right] \frac{M_{\rm Pl}^2}{\mu^4} \,,
\end{align}
which is typically very large, $k\gg1$.
This reflects the fact that $X$ is
stabilized by the strong dynamics close to the origin,
$\left<X\right> \ll M_{\rm Pl}$.
$H_J$ is the Hubble rate in the Jordan frame,
while $m_{1\ell}^J \simeq 0.02\,\lambda^2\Lambda$ denotes
the effective Polonyi mass in the IYIT model.
This mass is generated via meson loops at one-loop level~\cite{Chacko:1998si}.
The Polonyi field is stabilized at $\left<X\right>$ as given
in Eq.~\eqref{eq:XVEV} only as long as $\left<X\right>$  does not induce
masses for the IYIT quarks larger than $\Lambda_0$.
This defines a critical field value, $X_c \simeq \sqrt{2}/\lambda\,\Lambda_0$,
above which the Polonyi potential changes from a quadratic to
a logarithmic form.
The requirement that $\left<X\right>\lesssim X_c$ then translates into a lower
bound, $\lambda_{\rm min}\left(\Lambda,\left<\rho\right>\right) \sim 0.1\cdots1$,
on $\lambda$.
At the same time, we impose an upper bound, $\lambda \lesssim \lambda_{\rm pert} \simeq 4$,
so that non-calculable higher-dimensional terms in the K\"ahler potential,
which scale like $\lambda^2/\left(16\pi^2\right)$~\cite{Chacko:1998si}, are
suppressed by at least half an order of magnitude.
In fact, for definiteness, we will set
$\lambda^2 = \lambda_{\rm min}\,\lambda_{\rm pert}$ in the following.
Meanwhile, a given value of $\lambda$ implies a lower bound on $\rho$,
such that neither of $\lambda_\pm$ becomes larger than $\lambda_{\rm pert}$.
For our choice of $\lambda$, we find
$\rho^{-2} < \rho_{\rm pert}^{-2} = \left(\lambda_{\rm min}/\lambda_{\rm pert}
+\lambda_{\rm pert}/\lambda_{\rm min}\right)/2$,
where $\rho_{\rm pert}$ depends on the dynamical scale $\Lambda$ via $\lambda_{\rm min}$.


The Polonyi field is a linear combination of the charge
eigenstates $Z_\pm$.
It hence enters into the D-term potential,
where it mixes with the field $Y = \left(Z_+-Z_-\right)/\sqrt{2}$.
This mixing destabilizes the vacuum in Eq.~\eqref{eq:XVEV},
unless $\left|\Delta m_{XY}\right|^2 = \left|g^2q_0^2\,\xi\right| < m_{1\ell}^J\,m_F$,
which translates into an upper bound on $g$.
In our analysis of the inflationary dynamics, we will evaluate the bounds
on $\rho$ and $g$ numerically.
The lesson from these two bounds is the following:
$\rho > \rho_{\rm pert}$ guarantees that we can
neglect nonperturbative corrections to the K\"ahler potential,
while $g < g_{\rm max}$ ensures a stable vacuum in the SUSY-breaking sector.


The constant $w$ in Eq.~\eqref{eq:Polonyi} needs to be tuned to
\begin{align}
w_0 = \left(1-\frac{4}{3k}\right)^{1/2} \frac{\mu^2 M_{\rm Pl}}{\sqrt{3}} \,,
\end{align}
so as to reach zero cosmological constant in the SUSY-breaking vacuum.
We then obtain for $m_{3/2} = \left<W\right>/M_{\rm Pl}^2$,
\begin{align}
m_{3/2} = 
\left(1-\frac{4}{3k}\right)^{1/2}
\frac{\left(1-f\right)k +2/3}{\left(1-f\right)k-4/3}
\frac{\mu^2}{\sqrt{3} M_{\rm Pl}} \,.
\end{align}
Making use of our choices for the parameters $\lambda$ and $\rho$,
we arrive at the following phenomenological relation,
\begin{align}
\frac{m_{3/2}}{10^{11}\,\textrm{GeV}} \sim
\left(\frac{\sqrt{\xi}}{10^{15}\,\textrm{GeV}}\right)^{5/2} \sim
\left(\frac{\Lambda}{10^{15}\,\textrm{GeV}}\right)^{5/2} \,,
\label{eq:m32xi}
\end{align}
which illustrates that SUSY is broken at a high scale.
All mass scales in the IYIT sector are now solely controlled by $\Lambda$. 
This scale is dynamically generated via dimensional transmutation,
meaning that our model does not require any hard dimensionful input scale.
Eq.~\eqref{eq:m32xi} sets the stage for our particular
implementation of DHI.


We consider the following contributions to the kinetic function
and superpotential from the inflationary sector,
\begin{align}
F_\text{inf} & = \frac{\chi}{2} (S^\dagger+S)^2 - \frac{1}{2}(1-\chi)(S^\dagger-S)^2
+ \Phi^\dagger \Phi + \bar \Phi^\dagger \bar \Phi \,, \nonumber \\
W_\text{inf} & = \kappa\, S\, \Phi \bar \Phi \,. 
\label{eq:setupINF}
\end{align}
Taking $|\chi| \ll 1$, the real scalar component $\sigma$ of $S$ acquires an
approximate shift symmetry and will play the role of the inflaton.
The so-called waterfall fields $\Phi, \bar \Phi$ carry $B$$-$$L$ charges $\pm q$.
Due to their $\sigma$-dependent mass spectrum, one of the scalar waterfall
degrees of freedom becomes tachyonic at $\sigma \leq \sigma_c$,
leading to a phase transition that ends inflation.
The sequestered structure of Eq.~\eqref{eq:F} protects the waterfall
fields from acquiring soft SUGRA masses of the order of $m_{3/2}$,
which could prevent this phase transition.
Note that since $B$$-$$L$ is already broken by
$\langle M_\pm \rangle \neq 0$ \textit{during} inflation,
the production of cosmic strings at the end of inflation
can be avoided~\cite{Domcke:2014zqa}.


The scalar inflaton potential at $\sigma \gg \sigma_c$ is given by
\begin{align}
V(\sigma) \simeq {\cal C}^4(\sigma) \left[V_{D}^J + V_F^J(\sigma) 
+ \frac{Q_J^4(\sigma)}{16 \,\pi^2} \ln \left(x(\sigma)\right)\right] \,,
\label{eq:Vinf}
\end{align}
with the tree-level D- and F-term contributions being
\begin{align}
V_{D}^J = \frac{1}{2} \, q_0^2 \,  g^2  \, \xi^2  \,, \quad
V_F^J(\sigma) = \frac{- f(\sigma)}{1 - f(\sigma)} \, \mu^4 \,.
\end{align}
Similarly as in global SUSY, inflation is driven by the constant
D-term potential induced by the FI term~\eqref{eq:xi}.
$V_F^J$ arises due to F-term SUSY breaking in the
hidden sector and vanishes in the true vacuum at $\sigma = 0$.
The function $f = (1 - 2 \chi)\, \chi \sigma^2/(3 M_\text{Pl}^2) \ll 1$
is introduced below Eq.~\eqref{eq:XVEV}. The third term in Eq.~\eqref{eq:Vinf}
is the effective one-loop potential arising from integrating out the waterfall multiplets.
Here, the renormalization scale $Q_J^4 = q^2 m_D^4 + \delta m^4$
and $x = (m_{\rm eff}^2 + 2 H_J^2)/Q_J^2$ are given in terms
of the various contributions to the masses of the waterfall fields:
$m_D^2 = g^2q_0\xi$ is induced by the D term, $m_{\rm eff}^2 = \kappa^2 \sigma^2/2$
follows from the coupling in $W_{\rm inf}$, and 
$\delta m^2 \simeq m_{3/2}\,m_{\rm eff}$ is a bilinear soft mass in consequence
of $R$ symmetry breaking.
The critical inflaton field value $\sigma_c$ is obtained once
$m_{\rm eff}^2 + 2\,H_J^2 - Q_J^2 = 0$, which results in 
$\sigma_c^2 = 2/\kappa^2\left(Q_J^2-2\,H_J^2\right)$.
The conformal factor ${\cal C}^2 = - 3 M_\text{Pl}^2/\Omega_\text{tot}$
translates the Jordan-frame potentials to their counterparts in the Einstein frame.


Integrating out the heavy $B$$-$$L$ gauge fields results in
additional gauge-mediated soft masses, $m_{\rm gm}$, for the waterfall
fields~\cite{Intriligator:2010be} (see also \cite{Babu:2015xba}).
However, at the level of the effective one-loop potential for the inflaton,
these radiative  corrections merely represent a two-loop effect.
The inflaton itself, being a gauge singlet, receives by contrast no
gauge-mediated soft mass.
Moreover, one can show that,
in the parameter range of interest, $m_{\rm gm}$ is always
outweighed by the tree-level mass induced by the D-term 
potential, $m_{\rm gm} \lesssim 0.03\: m_D$.
For these reasons, we will neglect the effect
of gauge mediation in the following.

\begin{figure}
\begin{center}
\includegraphics[width=0.45\textwidth]{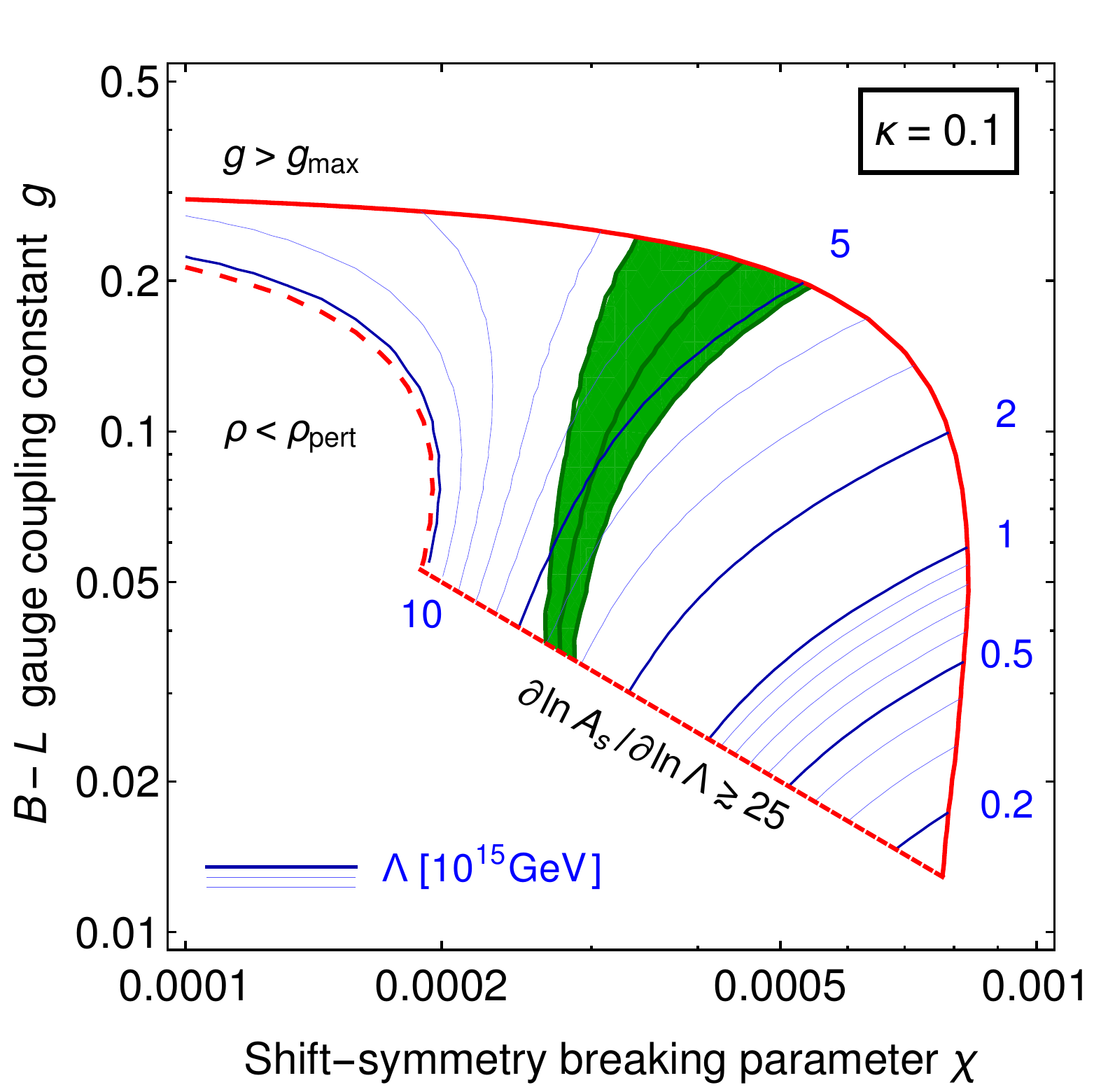}
\caption{Theoretical and experimental constraints.
The green band indicates values of the spectral index
in agreement (at $2\,\sigma$) with the Planck 2015 data,
$n_s = 0.9677 \pm 0.006$~\cite{Ade:2015lrj}.
The blue contours indicate the values of the dynamical
scale $\Lambda$ in the IYIT sector, such that $A_s = A_s^{\rm obs}$.
The red lines are boundaries of the parameter space due
to the requirement of perturbativity ($\rho > \rho_\text{pert}$)
and a stable vacuum ($g < g_\text{max}$) in the SUSY-breaking
sector as well as not too large fine-tuning in the parameters
of the inflation sector ($\partial \ln A_s/\partial \ln \Lambda \lesssim 25$).}
\label{fig:inflation}
\end{center}
\end{figure}

We solve the slow-roll equation of motion numerically,
\begin{equation}
K_{S^\dagger S}(\sigma) \, V(\sigma) \, \sigma'(N) = M_\text{Pl}^2 \, V'(\sigma)\,,
\end{equation}
to obtain the predictions for the CMB observables at $N_* \simeq 55$ e-folds before the end of inflation. With
\begin{equation}
\varepsilon = \frac{M_\text{Pl}^2}{2} \left( \frac{V'(\hat \sigma)}{V} \right)^2 \,, \quad \eta = M_\text{Pl}^2 \frac{V''(\hat \sigma)}{V} \,,
\end{equation}
where derivatives with respect to the canonically normalized field $\hat \sigma$ can be obtained by $\partial \hat \sigma/\partial \sigma = \sqrt{K_{S^\dagger S}}$, the amplitude of the scalar perturbation spectrum, its tilt and the tensor-to-scalar ratio are obtained as
\begin{equation}
A_s = \frac{V}{24 \pi^2 \varepsilon M_P^4} \,, \quad n_s = 1 - 6 \varepsilon + 2 \eta \,, \quad r = 16 \, \varepsilon \,,
\end{equation}
evaluated at $\sigma(N_*)$.
Requiring $A_s = A_s^{\rm obs} = 2.1 \times 10^{-9}$~\cite{Ade:2015lrj} fixes
$\Lambda$ (or equivalently $\xi$). 
The parameter $\kappa \neq 0$ explicitly breaks the shift symmetry in the superpotential,
which leads us to expect that $\kappa \lesssim 1$. On the other hand,
for $\kappa \ll 1$, the correct spectral index can only be obtained if
the SUGRA contributions become much larger than the one-loop
contributions~\cite{Domcke:2017rzu}. We thus set $\kappa = 0.1$.
In this regime, inflation occurs at field values slightly below
the Planck scale. For simplicity, we also fix the $B$$-$$L$ charges to $q = 2 q_0 = -2$,
inspired by neutrino mass generation (see below). We depict our results
in the remaining $(\chi, g)$ plane in Fig.~\ref{fig:inflation}. 
$r$ is of $\mathcal{O}\left(10^{-6}\cdots10^{-4}\right)$,
which is, similarly as in FHI, far below current bounds.


These results are very well reproduced by approximate analytical expressions
for the slow-roll parameters,
\begin{align}
\varepsilon & \simeq
\left(\frac{M_{\rm Pl}}{\sigma/\sqrt{2}}\right)^2 \left[
\left(1 +\delta_\varepsilon^4\right)\frac{q^2g^2}{16\pi^2}\frac{D_0^2}{V_{D}^J}
- f \frac{F_0^2}{V_{D}^J}
 \right]^2 , \label{eq:epsfinal} \\
- \eta & \simeq  \left(\frac{M_{\rm Pl}}{\sigma/\sqrt{2}}\right)^2 \left[
\left(1-\delta_\eta^4\right)\frac{q^2g^2}{16\pi^2}\frac{D_0^2}{V_{D}^J}
+ f\, \frac{F_0^2}{V_{D}^J} \right] , \label{eq:etafinal}
\end{align}
with $F_0^2 = \mu^4 \gtrsim D_0^2 = q_0^2 g^2 \xi^2$ and
\begin{align}
(\delta_\varepsilon/\delta)^4 = \ln x+\frac{1}{2}  \,, \:\:
(\delta_\eta/\delta)^4  = \ln x+\frac{1}{2} + \frac{2+\delta^4}{1+\delta^4} \,. 
\label{eq:deltas}
\end{align}
Here, $\delta^2 = \delta m^2/\left(q\, m_D^2\right)$ parametrizes the effect of the soft B-term mass, $\delta m$, in the one-loop potential. Lowering the spectral index compared to DHI in global SUSY becomes possible due to the negative mass-squared induced by the SUGRA contributions to the tree-level F-term potential, reflected by the last term in Eq.~\eqref{eq:etafinal},
\begin{align}
\Delta \eta \simeq - \frac{2 \chi}{3} \left(\frac{m_{3/2}}{H_J} \right)^2 \,, \quad
H_J^2 \simeq \frac{V_D^J}{3 M_\text{Pl}^2} \,.
\end{align}
Successful inflation is thus due to the interplay of the one-loop contribution and the SUGRA-induced mass, with the latter being suppressed by an approximate shift symmetry, $\chi \sim 10^{-4}$.
We note that $F_{\rm inf} \supset \chi\,\sigma^2$
in Eq.~\eqref{eq:setupINF} might, e.g., arise
from further shift-symmetry breaking terms in the superpotential.
Suppose the inflaton couples to superheavy multiplets with strength $\kappa'$.
Integrating out these fields results in an effective
K\"ahler potential, $K_{1 \ell} \sim \kappa'^2/(16 \pi^2) \, |S|^2$~\cite{Gaillard:1993es}.
With $\kappa' \sim \kappa \sim 0.1$, this is of just the right order
to explain the required value of $\chi$.
%


In the viable region of parameter space, inflation occurs either
near a hill-top (i.e., a local maximum in the scalar potential)
or near an inflection point, depending on the exact
values of $\chi$ and $g$~\cite{Domcke:2017rzu}. 
The hill-top regime may suffer from an initial conditions problem.
For particular parameter values, there is however a false vacuum
at large field values.
From there, $\sigma$ could tunnel to the correct
side of the hill-top, thereby setting off inflation in our Universe. 
The inflection-point regime allows, by contrast, to start out
at super-Planckian field values.


In FHI, $n_s \simeq 0.96$ is obtained
from the interplay of the one-loop potential and the
SUGRA tadpole, which is \textit{linear} in the inflaton field.
The tadpole also renders the question of initial conditions more subtle~\cite{Buchmuller:2000zm}.
Its size is controlled by $m_{3/2}$, an independent parameter, 
which can be chosen in accord with low-scale SUSY breaking.


Let us conclude.
We have presented a complete and phenomenologically viable SUGRA model of DHI,
in which inflation is driven by the D term of a gauged $U(1)_{B-L}$ symmetry.
Our model unifies the dynamics of dynamical SUSY breaking in the hidden sector,
DHI, and spontaneous $B$$-$$L$ breaking.
It links all relevant energy scales to the dynamical scale in the hidden sector,
the magnitude of which is fixed by the amplitude of the CMB power 
spectrum, $\Lambda \simeq 5\times10^{15}\,\textrm{GeV}$.
This value is remarkably close to the GUT scale,
$\Lambda_{\rm GUT} \sim 10^{16}\,\textrm{GeV}$.


We based our construction on three assumptions:
dynamical SUSY breaking as the origin of the FI term, Jordan-frame supergravity
with canonical kinetic terms, and an approximate inflaton shift symmetry.
These assumptions remedy all shortcomings
of standard DHI:
Our FI term is a field-dependent FI term.
The associated modulus is stabilized via F-term SUSY breaking.
The MSSM sfermions are stabilized against tachyonic D-term-induced
masses thanks to their direct coupling to the hidden sector in the kinetic
function $F_{\rm tot}$ \eqref{eq:F}.
Owing to our Jordan-frame description, the inflaton sector sequesters from
the hidden sector, such that the fields in the inflation
sector pick up no dangerous gravity-mediated soft masses.
Meanwhile, a slight breaking of the shift symmetry provides
a small SUGRA correction to the inflaton potential, $V\supset -\chi\, m_{3/2}^2\, \sigma^2$,
that allows to reproduce $n_s \simeq 0.96$.
As $B$$-$$L$ is spontaneously broken in the hidden sector
already during inflation, no dangerous cosmic strings are produced during
the waterfall transition.


Our model has important phenomenological consequences.
For instance, if we assign $B$$-$$L$ charge $q = -2$ to the waterfall field $\Phi$,
it can couple to the right-handed neutrinos $N_i$ in the seesaw
extension of the MSSM, $W \supset h_{ij}/2\,\Phi\, N_i N_j$.
For $q_0\,\xi <0$, it is the field $\Phi$ that acquires a nonzero VEV
during the waterfall transition, $\left<\Phi\right> = \left|q_0/q\,\xi\right|^{1/2}$,
whereas $\left<\bar{\Phi}\right>$ remains zero.
This VEV generates the Majorana mass matrix for the right-handed
neutrinos, $M_{ij} = h_{ij} \left<\Phi\right>$, and, hence, sets the stage
for the seesaw mechanism and leptogenesis~\cite{Buchmuller:2010yy}.
Besides, our model predicts a superheavy SUSY mass spectrum.
Only the lightest neutralino may have a fine-tuned small
mass, so as to evade overproduction in gravitino
decays~\cite{Moroi:1999zb}.
Together with the corresponding chargino, this neutralino is then expected
to be the only superparticle at low energies.
It constitutes thermal neutralino dark matter and can be
searched for in direct detection experiments.%
\footnote{In our model, the mediation of spontaneous SUSY breaking 
to the visible sector is essentially described by the framework of
pure gravity mediation (PGM)~\cite{Ibe:2006de}.
In this mediation scheme, one is, e.g., able to achieve a light wino mass
by tuning higgsino threshold corrections
against the usual gaugino masses from anomaly mediation~\cite{Ibe:2012hu}.
For $\tan\beta = v_u / v_d \simeq 1$ and Higgs mass parameters of the order of
the gravity mass, $\left|\mu_H\right| \simeq \left|B\right| \simeq m_{3/2}$, PGM readily allows
to push the wino mass down to $m_{\rm wino} \simeq 2.7\,\textrm{TeV}$, so that it may
constitute ordinary thermal dark matter in the form of weakly interacting massive
particles (WIMPs).
This fine-tuning might be the result of anthropic selection.
In addition, the wino mass may also receive threshold and anomaly-mediated
corrections from heavy vector matter multiplets charged under $SU(2)_L$ and/or
threshold corrections from the F terms of
flat directions in KSVZ-type axion models~\cite{Harigaya:2013asa}.
In this case, these contributions would also play a role in tuning the wino mass.}
One may also hope to probe our model in gravitational-wave
(GW) experiments.
Depending on further model assumptions, the $B$$-$$L$ phase transition
may give rise to observable signals~\cite{Maggiore:1999vm}.
Likewise, if the shift symmetry is realized along the imaginary component of
$S$, the inflaton may have an axion-like coupling to gauge fields,
$\mathcal{L}_{\rm eff} \supset \text{Im}(S)\, F\tilde{F}$.
This could drastically enhance the GW signal
from inflation~\cite{Cook:2011hg}. 


Our model leaves open several questions that call for further
exploration:
For instance, one may ask what UV physics underlies the kinetic
function in Eq.~\eqref{eq:F}.
It would be interesting to derive this structure
from the viewpoint of a higher-dimensional brane-world
scenario or from a strongly coupled conformal field theory~\cite{Inoue:1991rk}.
We have only briefly sketched the mechanism
of sfermion mass generation.
It would, therefore, be desirable to devise a model that accounts for
the origin of the scale $M_*$ in Eq.~\eqref{eq:F}.
These questions are however beyond the scope of this work.
We conclude by stressing that our dynamical SUGRA model 
resolves all issues of standard DHI.


The authors wish to thank T.~T.~Yanagida
for inspiring discussions at Kavli IPMU at the University of Tokyo
in the early stages of this project in 2014.
The authors are grateful to B.~v.~Harling and L.~Witkowski for helpful remarks
and thank W.~Buchm\"uller,
E.~Kiritsis and T.~T.~Yanagida for useful comments on the manuscript.
V.\,D.\ acknowledges financial support from the UnivEarthS Labex program at
Sorbonne Paris Cit\'e (ANR-10-LABX-0023 and ANR-11-IDEX-0005-02) and
the Paris Centre for Cosmological Physics.
This project has received support from the European Union's Horizon 2020
research and innovation programme under the Marie Sk\l odowska-Curie grant
agreement No.\ 674896 (K.\,S.).
K.\,S.\ acknowledges the hospitality of the APC/PCCP group
during a stay at Paris Diderot University, where this project was finished.



\end{document}